\documentstyle[prd,aps,12pt]{revtex}
\begin{document} 

\tighten
\draft
\preprint{DAMTP96-42} 

\title{Generalised Heisenberg Relations} 

\author{
Dorje C. Brody 
and 
Lane P. Hughston 
} 
\address{* Department of Applied Mathematics and 
Theoretical Physics, \\ University of Cambridge,  
Silver Street, Cambridge CB3 9EW U.K.} 
\address{$\dagger$ Merrill Lynch International, 
25 Ropemaker Street, London EC2Y 9LY U.K. \\ 
and King's College London, The Strand, London 
WC2R 2LS, U.K.} 

\date{\today} 

\maketitle 

\begin{abstract} 
A geometric framework for quantum statistical estimation 
is used to establish a series of higher order corrections 
to the Heisenberg uncertainty relations associated with 
pairs of canonically conjugate variables. \par 
\end{abstract} 



An efficient geometric approach to parameter estimation, based 
upon Hilbert space geometry, has been developed recently 
\cite{dblh}. The idea is to regard a statistical model as being 
embedded into a real Hilbert space ${\cal H}$, thereby allowing 
statistical estimation theory to be expressed neatly in terms of 
the geometry of ${\cal H}$. In particular, the concept of a 
statistical model can also be formulated in the case of a quantum 
mechanical state space, after a compatible complex structure is 
introduced on the underlying real Hilbert space, thus leading to 
a succinct characterisation of the quantum estimation problem.\par 

In the present Letter, this geometrical framework is applied to 
establish a series of corrections to the Heisenberg uncertainty 
relations. For example, in the case of the measurement of a 
`position' parameter $q$, for which an unbiased estimator is 
given by an operator $Q$, with conjugate momentum operator $P$, 
where $i[P,Q]=1$ ($\hbar=1$), we find an infinite set of 
corrections to the uncertainty lower bound, given in terms of 
higher order central moments of the distribution of $P$. If we 
write $\mu_{n} = \langle(P-\langle P\rangle)^{n}\rangle$, 
these corrections take the form 
\begin{eqnarray} 
\Delta Q^{2}\Delta P^{2}\ &\geq&\ \frac{1}{4} \left( 1 + 
\frac{(\mu_{4}-3\mu_{2}^{2})^{2}}{\mu_{6}\mu_{2}-\mu_{4}^{2}} 
\right. \nonumber \\ 
& &\ + \left. 
\frac{\mu_{2}[\mu_{8}(\mu_{4}-3\mu_{2}^{2})+\mu_{6}(8\mu_{4}\mu_{2} 
- \mu_{6}) - 5\mu_{4}^{3}]^{2}}{(\mu_{10}(\mu_{6}\mu_{2}- 
\mu_{4}^{2}) + 2\mu_{8}\mu_{6}\mu_{4} - \mu_{8}^{2}\mu_{2} - 
\mu_{6}^{3})(\mu_{6}\mu_{2}-\mu_{4}^{2})} +\ \cdots \right) \ . 
\label{eq:1} 
\end{eqnarray} 
The statistical model ${\cal M}$ in this case is the one-parameter 
family of states $|\psi_{q}\rangle$ obtained by the action of the 
unitary operator $e^{iqP}$ on an initial state $|\psi_{0}\rangle$. 
The problem is to use measurements of the observable $Q$ to 
estimate the value of $q$, provided that the actual state of the 
system lies somewhere on the manifold ${\cal M}$. \par 

In order to pursue these ideas further, first we review a few of 
the basic concepts of statistical geometry. Consider a real Hilbert 
space ${\cal H}$, with a symmetric inner product denoted $g_{ab}$, 
and write $\xi^{a}$ for a typical vector 
in ${\cal H}$. A random variable in ${\cal H}$ is represented by a 
symmetric operator $X_{ab}$, with expectation $E_{\xi}[X] = 
X_{ab}\xi^{a}\xi^{b}/ \xi^{c}\xi_{c}$ in the state $\xi^{a}$. The 
variance of $X_{ab}$ is then 
\begin{equation} 
{\rm Var}_{\xi}[X]\ =\ \frac{\tilde{X}_{ab}\tilde{X}^{b}_{c} 
\xi^{a}\xi^{c}}{g_{de}\xi^{d}\xi^{e}}\ , 
\end{equation} 
where $\tilde{X}_{ab} = X_{ab} - E_{\xi}[X]g_{ab}$ represents the 
deviation of $X_{ab}$ from its mean. In these expressions $\xi^{a}$ 
is required to lie in the domain of $X_{ab}$, which in the case of 
a bounded operator is the whole of ${\cal H}$. If $\xi^{a}$ is 
normalised, we have $g_{ab}\xi^{a}\xi^{b}=1$, whence $\xi^{a}$ is a 
point on the unit sphere ${\cal S}$ in ${\cal H}$. By a 
`statistical model', we mean a submanifold ${\cal M}$ of ${\cal S}$, 
given parametrically by $\xi^{a}(\theta^{i})$ where $\theta^{i}$ 
are local coordinates. The Riemannian metric induced on ${\cal M}$, 
known as the Fisher-Rao metric, is given by \cite{dblh,rao} 
\begin{equation} 
{\cal G}_{ij}\ =\ 4 g_{ab}\partial_{i}\xi^{a} 
\partial_{j}\xi^{b} \ , 
\end{equation} 
where $\partial_{i}=\partial/\partial\theta^{i}$ (the factor 
of 4 is conventional). For simplicity, 
in what follows we consider the case where ${\cal M}$ is 
one-dimensional, and for which the Fisher information is thus 
${\cal G} = 4g_{ab}\dot{\xi}^{a}\dot{\xi}^{b}$, where the dot 
denotes $\partial/\partial\theta$.\par 

Now we are able to pose the following statistical estimation 
problem. Given the measurement of an observable $X_{ab}$, we are 
interested in a one-parameter family $\xi^{a}(\theta)$ of possible 
states characterising the distribution of the outcome. The 
parameter $\theta$, which is understood to lie in some specified 
range, determines the unknown state of nature, and 
we wish to estimate $\theta$ with the given data. For such an 
estimation problem a lower bound can be established 
for the variance of the estimate. Comparing the variance of our 
estimate to the lower bound, we can enquire to what 
extent the estimator is efficient. An elementary geometrical 
derivation of this bound is as follows. Given a curve 
$\xi^{a}(\theta)$ in $\cal{S}$, we say that a random 
variable $T_{ab}$ is an {\it unbiased estimator} for a function 
$\tau(\theta)$ if 
\begin{equation} 
\frac{T_{ab}\xi^{a}\xi^{b}}
{g_{cd}\xi^{c}\xi^{d}}\ =\ \tau(\theta)\ , 
\end{equation} 
along the given trajectory. 
Since $\tau$, regarded as a function of $\xi^{a}$, is homogeneous, 
we can extend its definition to the whole of the domain of $T_{ab}$ 
in ${\cal H}$, and thus define the gradient $\nabla_{a}\tau$, where 
$\nabla_{a} = \partial/\partial\xi^{a}$. A short calculation shows 
that $\nabla_{a}\tau=2{\tilde T}_{ab}\xi^{b}/g_{cd}\xi^{c}\xi^{d}$, 
where ${\tilde T}_{ab}=T_{ab}-\tau g_{ab}$, and as a consequence 
we deduce that the variance of the estimator $T$ is given by 
\begin{equation} 
{\rm Var}_{\xi}[T]\ =\ 
\frac{1}{4}g^{ab}\nabla_{a}\tau\nabla_{b}\tau 
\label{eq:VAR} 
\end{equation} 
for any state on the given curve in ${\cal S}$. This formula gives us 
a useful geometrical interpretation for the variance of an estimator. 
\par 

The squared length of the gradient vector $\nabla_{a}\tau$ 
can be expressed as the sum of the squares of its orthogonal 
components with respect to a suitable basis. To this end, we choose 
a set of vectors comprising the state $\xi^{a}$ and 
its higher order derivatives. Letting $\hat{\xi}^{(n)a}$ denote the 
component of $\xi^{(n)a}=d^{n}\xi^{a}/d\theta^{n}$ orthogonal to both 
$\xi^{a}$ and all of its derivatives of order less than $n$, we 
obtain a  set of orthonormal vectors is given 
by $\{ \hat{\xi}^{(n)a}/\sqrt{\hat{\xi}^{(n)b}\hat{\xi}^{(n)}_{b}}\}$ 
$(n=0,1,2,\cdots)$. It follows from the basic relation 
(\ref{eq:VAR}) that 
\begin{equation} 
{\rm Var}_{\xi}[T]\ \geq\ \frac{1}{4} 
\sum_{n} \frac{(\hat{\xi}^{(n)a}\nabla_{a}\tau)^{2}}
{\hat{\xi}^{(n)b}\hat{\xi}^{(n)}_{b}}\ . \label{eq:gbb} 
\end{equation} 
Since each term in the sum on the right hand side of (\ref{eq:gbb}) 
is positive, we obtain a set of inequalities, which we 
refer to as generalised Bhattacharyya variance 
lower bounds \cite{dblh}. An essential difference here from the 
corresponding classical inequalities \cite{bh} is that the bounds 
in (\ref{eq:gbb}) are not necessarily independent of the specific 
choice of estimator. In our approach to the quantum estimation 
problem below, however, we find a set of bounds that are 
systematically independent of the estimator $T$. \par 

We shall consider the case where ${\cal M}$ is the submanifold of 
a quantum mechanical state space generated by the action of a 
one-parameter family of unitary transformations starting from some 
given initial state. In particular, we shall examine 
the case of state vector $\xi^{a}(t)$ that satisfies the 
Schr\"odinger equation: 
\begin{equation} 
\dot{\xi}^{a}\ =\ J^{a}_{\ b}{\tilde H}^{b}_{\ c}\xi^{c}\ . 
\label{eq:sch} 
\end{equation} 
Here, we identify $\theta$ in the analysis above with the time 
parameter $t$, and the tensor $J^{a}_{\ b}$, satisfying 
$J^{a}_{\ c}J^{c}_{\ b} = -\delta^{a}_{\ b}$, is the usual complex 
structure on ${\cal H}$. The `mean-adjusted' Hamiltonian 
${\tilde H}_{ab}$ is symmetric and Hermitian, and describes the 
deviation of the Hamiltonian from its mean in the given state. 
The variance of the Hamiltonian is then given by 
$\Delta H^{2} = g_{ab}{\dot \xi}^{a}{\dot \xi}^{b}$. \par 

Note that in our approach the Hilbert space of quantum 
mechanics is treated as a real Hilbert space ${\cal H}$ endowed 
with a metric $g_{ab}$ and a compatible complex structure 
$J^{a}_{b}$. This is entirely equivalent to the conventional 
formulation of quantum mechanics based on complex Hilbert 
space \cite{dblh2}, but has the virtue of being more geometrically 
transparent in such a way, in particular, as to make the link 
with statistical geometry much clearer. \par 

Now, suppose that we have a large number of independent identical 
systems, each of which evolves from an initial state $\xi^{a}(0)$ 
under the influence of the Hamiltonian $H_{ab}$. The problem is to 
estimate how much time $t$ has elapsed since the initial 
preparation. Let $T_{ab}$ be an unbiased estimator for the time 
parameter in this problem, so 
\begin{equation} 
\frac{T_{ab}\xi^{a}\xi^{b}}{g_{cd}\xi^{c}\xi^{d}}\ 
=\ t \label{eq:time} 
\end{equation} 
along the trajectory $\xi^{a}(t)$. The existence of such an 
estimator is demonstrated, for example, in \cite{hol}. As a 
consequence of (\ref{eq:sch}) and (\ref{eq:time}), it follows 
that $2T_{ab}J^{b}_{c}H^{c}_{d}\xi^{a}\xi^{d} = 1$ along the 
normalised trajectory $\xi^{a}(t)$, which can be viewed as a `weak' 
form of the canonical commutation relation. In other words, for an 
estimator, as opposed to an observable, we merely require an operator 
that satisfies the canonical commutation relation in expectation 
along the designated trajectory with reference to which the 
relevant measurement operation is being performed. \par 

We are interested in finding an explicit form of the variance 
lower bounds (\ref{eq:gbb}) for quantum mechanical situations. 
When $\xi^{a}$ satisfies the Schr\"odinger equation (\ref{eq:sch}), 
we find that our series of orthogonal vectors is given by 
\[ 
\left\{ \xi^{a},\ \ \dot{\xi}^{a},\ \ \ddot{\xi}^{a}-(\ddot{\xi}^{b}
\xi_{b})\xi^{a},\ \ \stackrel{...\ }{\xi^{a}} - 
\frac{\stackrel{...\ }{\xi^{b}}\dot{\xi}_{b}}
{\dot{\xi}^{c}\dot{\xi}_{c}}\dot{\xi}^{a}, \ \cdots \right\} \ . 
\] 
Let $\{\Psi^{a}_{n}\}$ denote this series of vectors, so $\Psi^{a}_{0} 
= \xi^{a}$, $\Psi^{a}_{1}=\dot{\xi}^{a}$, and so on. We note that  
$\nabla_{a}t = 2{\tilde T}_{ab}\xi^{b}$ for normalised states in the 
domain of $T_{ab}$, so we can rewrite (\ref{eq:gbb}) in the form  
\begin{equation} 
\Delta T^{2}\ \geq\ \frac{1}{4}\sum_{n} 
\frac{(2\Psi^{a}_{n}{\tilde T}_{ab}\xi^{b})^{2}}
{g_{ab}\Psi^{a}_{n}\Psi^{b}_{n}} \ . \label{eq:gbb2}
\end{equation} 
Since $\Psi^{a}_{n}$ can be expressed in terms of linear 
combinations of the derivatives of $\xi^{a}$, we find that for 
odd integers $n$, the following remarkable identity can be used 
to simplify equation (\ref{eq:gbb2}): 
\begin{equation} 
2T_{ab}\xi^{(n)a}\xi^{b}\ =\ (-1)^{m} n g_{ab} 
\xi^{(m)a}\xi^{(m)b}\ , \label{eq:ui}
\end{equation} 
where $m=\frac{1}{2}(n-1)$. Furthermore, a simple calculation 
shows that $g_{ab}\xi^{(m)a}\xi^{(m)b}=\mu_{2m}$, where $\mu_{2m}= 
\langle {\tilde H}^{2m}\rangle$ denotes the $2m$-th central moment 
of the Hamiltonian in the given state. Therefore, we find that for 
$n=0$ the 
contribution in (\ref{eq:gbb2}) vanishes, and the $n=1$ term gives 
$(4\mu_{2})^{-1}$, i.e., the Heisenberg lower bound. We also find 
\cite{dblh} that the $n=2$ term depends upon the choice of the 
estimator $T$, while for $n=3$ the result, which is independent of 
the choice of estimator, can be expressed in terms of the fourth 
cumulant of the distribution for the Hamiltonian $H$, as indicated 
in equation (1). We now proceed to deduce the general pattern of 
such higher order terms. \par 

In general, the even order contributions depend on the covariance 
of the estimator $T$ with $H^{2n}$, while by 
virtue of (\ref{eq:ui}) the odd order terms are manifestly 
independent of the specific choice of $T$. Hence we only consider 
odd order terms. Let $N_{n}=g_{ab}\Psi^{a}_{n}\Psi^{b}_{n}$ 
denote the denominator of the correction terms in (\ref{eq:gbb2}). 
Then, an exercise shows that 
\begin{equation} 
N_{n}\ =\ \frac{D_{2n}}{D_{2n-4}}\ , \label{eq:N(n)} 
\end{equation} 
where $D_{2n}$ is defined by the determinant 
\begin{equation} 
D_{2n}\ =\ 
\left| \begin{array}{clcc} 
\mu_{2n}   & \mu_{2n-2} & \cdots & \mu_{n+1} \\
\mu_{2n-2} & \mu_{2n-4} & \cdots & \mu_{n-1} \\
\vdots &      & \ddots & \vdots \\
\mu_{n+1}  & \mu_{n-1} & \cdots & \mu_{2} 
\end{array} \right|  \ ,  \label{eq:D} 
\end{equation}  
therefore, $D_{2}=\mu_{2}$, $D_{6}=\mu_{6}\mu_{2}-\mu_{4}^{2}$, and 
so on. We note, incidentally, that $N_{2}$ gives 
the curvature of the curve $\xi^{a}(t)$ in ${\cal S}$, and 
$N_{k}$ for $k>2$ determines the higher order `torsion' of $\xi^{a}$, 
in the sense of classical differential geometry. This can be seen 
by noticing that (\ref{eq:D}) is expressible in 
terms of the norm of the antisymmetric tensor obtained by 
skew-symmetrising the product of the following $m+1$ vectors: 
$\xi^{(n)}_{a}, \xi^{(n-2)}_{a}, \cdots, \xi^{(1)}_{a}$. 
For example, we have 
$D_{10}\ =\ 6\xi^{(5)}_{[a}\xi^{(3)}_{b}\xi^{(1)}_{c]} 
\xi^{(5)[a}\xi^{(3)b}\xi^{(1)c]}$, 
and so on. Standard statistical identities \cite{ks} guarantee 
$D_{2n}\geq0$. Thus, in order to 
find the general pattern for the higher order corrections, we 
must understand the structure of the numerators $2\Psi^{a}_{n} 
{\tilde T}_{ab}\xi^{b}$. Let us recall the way the basis 
vectors $\Psi^{a}_{n}$ are formed: 
\[ 
\left\{ \begin{array}{l} 
\Psi_{1}^{a}=\dot{\xi}^{a} \\ 
\Psi^{a}_{3}=\xi^{(3)a} - \frac{\xi^{(3)b}\Psi_{1b}}
{\Psi^{c}_{1}\Psi_{1c}}\Psi^{a}_{1} \\ 
\Psi^{a}_{5}=\xi^{(5)a} - \frac{\xi^{(5)b}\Psi_{3b}}
{\Psi^{c}_{3}\Psi_{3c}}\Psi^{a}_{3} - \frac{\xi^{(5)b}\Psi_{1b}}
{\Psi^{c}_{1}\Psi_{1c}}\Psi^{a}_{1} \\ 
\ \vdots 
\end{array} \right. 
\] 
That is, for each value of $n$, we subtract the components of 
$\Psi^{a}_{k}$ from $\xi^{(n)a}$ where $k<n$. Since we know 
$2\Psi^{a}_{1}{\tilde T}_{ab}\xi^{b}=1$, we obtain 
$2\Psi^{a}_{3}{\tilde T}_{ab}\xi^{b}=\mu_{2}^{-1}(\mu_{4}-
3\mu_{2}^{2})$ by use of (\ref{eq:ui}) and 
the above expression for $\Psi^{a}_{3}$. 
Furthermore, from these two results, we can then 
find $2\Psi^{a}_{5}{\tilde T}_{ab}\xi^{b}$, provided we know the 
coefficients of $\Psi^{a}_{k}$ $(k=1,3)$ in $\Psi^{a}_{5}$, and 
similarly for $\Psi^{a}_{7}$ and so on. 
In other words, once we find the pattern for these coefficients, 
we can express the resulting expression for 
higher order corrections in a recursive manner. \par 

We notice that the denominator terms in the coefficients are 
given by the expressions $N_{k}$ defined above.  
Hence we have reduced the problem to finding 
an expression for $\xi^{(n)a}\Psi_{ka}$ where $k<n$. If we define 
$F_{n,k}\equiv\xi^{(n)a}\Psi_{ka}/\Psi^{b}_{k}\Psi_{kb}$, then 
after some algebra we deduce that: 
\begin{eqnarray} 
F_{n,k}\ =\ \frac{(-1)^{\frac{n+k}{2}-1}}{D_{2k}}  
\left| \begin{array}{clcc} 
\mu_{n+k}   & \mu_{n+k-2} & \cdots & \mu_{n+1} \\
\mu_{2k-2} & \mu_{2k-4} & \cdots & \mu_{k-1} \\
\vdots &      & \ddots & \vdots \\
\mu_{k+1}  & \mu_{k-1} & \cdots & \mu_{2} 
\end{array} \right|  \ .  \label{eq:F} 
\end{eqnarray}  
In order to clarify the notation, we note that $F_{n,3}$ and 
$F_{n,1}$ are given respectively by: 
\[ 
F_{n,3}\ =\ (-1)^{m+1}\frac{1}{D_{6}} 
\left| \begin{array}{cc} 
\mu_{n+3} & \mu_{n+1} \\ 
\mu_{4} & \mu_{2} 
\end{array} \right| \ , 
\] 
\[ 
F_{n,1}\ =\ (-1)^{m}\frac{1}{D_{2}}\mu_{n+1} \ ,  
\] 
with $m=\frac{1}{2}(n-1)$. 
We now have obtained all the relevant identities needed 
in order to find a recursive formula for the higher order 
corrections. Let us define the numerator in (\ref{eq:gbb2}) 
by $U_{n}\equiv2\Psi^{a}_{n}{\tilde T}_{ab}\xi^{b}$. Then, from 
the following expression for $\Psi^{a}_{n}$:  
\begin{equation} 
\Psi^{a}_{n}\ =\ \xi^{(n)a} - \sum_{k=1,3,5,\cdots}^{n-2}  
F_{n,k}\Psi^{a}_{k}\ , 
\end{equation} 
a recursive formula for $U_{n}$ can be obtained in the form:  
\begin{equation} 
U_{n}\ =\ (-1)^{m} n \mu_{n-1} - \sum_{k=1,3,5,\cdots}^{n-2} 
F_{n,k}U_{k}\ , \label{eq:U} 
\end{equation} 
with $U_{1}=1$. Therefore, the  
uncertainty relation (\ref{eq:gbb2}) can be reexpressed as 
\begin{equation} 
\Delta T^{2}\Delta H^{2}\ \geq\ \frac{1}{4} 
\sum_{k=1,3,\cdots}\frac{\mu_{2}U_{k}^{2}}{N_{k}}\ . \label{eq:ghr} 
\end{equation} 
By input of the above relations (\ref{eq:U}) and (\ref{eq:ghr}), 
a suitable symbolic manipulation 
application can be made to produce expressions for the higher 
order terms purely in terms of the central moments 
$\mu_{2k}$ of the distribution of the Hamiltonian $H$. \par 

One might wonder whether the higher order contributions in 
(\ref{eq:ghr}) vanish, e.g., for Gaussian distributions. 
However, it turns out that the expressions for $U_{n}$ in 
(\ref{eq:U}) can be reexpressed in terms of linear combinations of 
the cumulants $\kappa_{2k}$ of the distribution for $H$. We recall 
that if $\phi(\lambda)$ is the characteristic function (Fourier 
transform) of the distribution for $H$, then the cumulants 
$\kappa_{r}$ are defined by the relation 
\begin{equation} 
\sum_{r}\kappa_{r}\frac{(i\lambda)^{r}}{r!}\ =\ 
\ln\phi(\lambda)\ , \label{eq:cum}
\end{equation} 
and for Gaussian distributions all the 
cumulants higher than the second order vanish. \par 

On the other hand, for quantum mechanical applications, the 
distribution of $H$ is not Gaussian, since $H$ is bounded below. 
Hence one might wish to know typical orders of magnitude 
for these corrections. To obtain a crude estimate of the 
sort of numbers that might arise, we consider a system where 
the distribution of the Hamiltonian can be modelled by 
a gamma distribution, for which the density function is of the form 
\begin{equation} 
p(H)\ =\ \frac{\sigma^{\gamma}}{\Gamma(\gamma)} 
e^{-\sigma H} H^{\gamma-1}\ , 
\end{equation} 
where $0\leq H\leq \infty$ and $\sigma, \gamma > 0$. In this 
case the resulting bound can be calculated, to second order, to be   
\begin{equation} 
\Delta T^{2}\Delta H^{2}\ \geq\ \frac{1}{4} 
\left( 1+\frac{18}{3\gamma^{2}+47\gamma+42}+\cdots \right) \ , 
\end{equation} 
with $\Delta H^{2}=\gamma/\sigma^{2}$. 
In particular, for $\gamma=1$, the distribution reduces to 
an exponential probability $p(H)=\sigma\exp(-\sigma H)$, and we 
find 
\begin{equation}  
\Delta T^{2}\Delta H^{2}\ \geq\ \frac{1}{4} 
\left( 1 + 0.196 + 0.063 + \cdots \right) \ ,  
\end{equation} 
to third order, where the terms on the right are independent of 
$\sigma$. These examples suggest an exponential decay for the 
higher order contributions, although in general this may 
not be the case. Extensive experimental data are available for 
the energy spectra of various systems, by use 
of which the associated higher order contributions can 
be evaluated. This would make for an interesting line of 
enquiry to pursue. One notices that in the case of an exponential 
distribution, the first two corrections already add up to a 
significant level of over $25\%$ of the lower bound. \par 

We observe that all the corrections in (\ref{eq:ghr}) are given in 
terms of even order central moments of the distribution for $H$. 
This, however, does not imply that odd moments do not exist for 
quantum trajectories. If the distribution for the conjugate 
variable is symmetric around its mean, then the odd moments 
vanish. However, as the 
situation stands, the system of orthogonal vectors 
$\{\Psi^{a}_{n}\}$ does not form a complete set of basis vectors. 
In particular, on the unit sphere ${\cal S}$, one can also consider 
the Cauchy-Riemann direction $\eta^{a}$ defined by $\eta^{a}\equiv 
J^{a}_{\ b}\xi^{b}$. By virtue of the Schr\"odinger equation 
(\ref{eq:sch}), $\eta^{a}$ is automatically orthogonal, for example, 
to $\xi^{a}$ and $\dot{\xi}^{a}$. By use of this direction 
field and its derivatives, we can complete the basis. The 
inclusion of such terms in the expansion for the variance 
inequality leads to further refined bounds \cite{dblh1}, which 
are not generally negligible if the distribution is asymmetric.\par 

Our approach here has been to take a geometric point of view in 
order to study the structure of the state space of quantum 
mechanics. The key to this approach is encapsulated in equation 
(\ref{eq:VAR}), where an essentially 
probabilistic, or statistical quantity, i.e., the variance, is 
expressed in terms of an essentially geometric quantity, the 
length of a vector. Presumably, the higher order terms in 
the uncertainty relation (\ref{eq:ghr}) can, in principle, 
be constructed by means of standard Fourier analysis. In that 
respect it would be interesting to compare the results we have 
obtained here to other approaches to generalised uncertainty 
relations \cite{ghr}. It should be emphasised, however, that the 
relations obtained therein take the form of `strong' dispersion 
bounds, in the sense that they apply to all states in the Hilbert 
space (or at any rate a dense subset thereof); whereas our results 
are much more refined in the sense that they apply in a specific, 
physical measurement theoretic context for the determination of 
a parameter value in the case of a single trajectory of 
states (cf. \cite{hol}). \par 

It is interesting to recall the fact that the higher order 
corrections can also be expressed by the cumulants of the 
distribution $p(H)$. This result suggests the possibility of 
constructing a cumulant expansion for the uncertainty 
relations. The cumulants can also be associated with 
connected graphs with fixed numbers of `legs', due to 
the logarithm in (\ref{eq:cum}), which can be compared with the 
use of Feynman diagrams for 
the free energy or ground state energy in quantum field 
theories. Therefore, it would be interesting to determine 
whether a neat graphic expansion for the uncertainty relations 
can be constructed. \par 

It should be pointed out that although the 
energy-time uncertainty is studied here as a matter of 
illustration, on account of its especially interesting 
features, analogous results hold for other conjugate variables, 
such as position and momentum, in which case the statistical 
manifold is given by the curve $d\xi^{a}/dq = 
J^{a}_{\ b}{\tilde P}^{b}_{\ c}\xi^{c}$ for some initial state 
$\xi^{a}(0)$. In this case both the momentum operator $P$ and 
the position operator $Q$, which acts as an estimator for $q$, 
are self adjoint. The arguments outlined above then lead to the 
inequality (\ref{eq:1}). It should be born in mind that the 
geometric formalism employed here, although highly effective, 
is not strictly essential, and the same results can be obtained 
by use of the familiar Dirac formalism, working in the 
Schr\"odinger representation. \par

The authors are grateful to C.M. Bender, A.P.A. Kent and 
B.K. Meister for useful discussions. DCB would like to thank 
PPARC for financial support. \par 


$*$ Electronic address: d.brody@damtp.cam.ac.uk \par 
$\dagger$ Electronic address: lane@ml.com \par

\end{document}